\documentclass[aps,prd,twocolumn,superscriptaddress,floatfix,nofootinbib]{revtex4}

\usepackage{color}
\usepackage[dvips]{graphicx}

\newcommand{\lsim}{\mathrel{\mathop{\kern 0pt \rlap
      {\raise.2ex\hbox{$<$}}}\lower.9ex\hbox{\kern-.190em $ \sim$}}}

\newcommand{\gsim}{\mathrel{\mathop{\kern 0pt
      \rlap{\raise.2ex\hbox{$>$}}}\lower.9ex\hbox{\kern-.190em $\sim$}}}

\newcommand{\beq}{\begin{equation}}
\newcommand{\eeq}{\end{equation}}

\newcommand{\be}{\begin{equation}}
\newcommand{\ee}{\end{equation}}
\newcommand{\beqarr}{\begin{eqnarray}}
\newcommand{\eeqarr}{\end{eqnarray}}

\begin{document}

\title{Impact of the recent results by the CMS and ATLAS
  Collaborations at the CERN Large Hadron Collider on an effective
  Minimal Supersymmetric extension of the Standard Model}

\thanks{Preprint number: DFTT 6/2011}


%
\author{S. Scopel}
\affiliation{Department of Physics, Sogang University\\
Seoul, Korea, 121-742}
\author{S. Choi} \affiliation{Department of Physics, Korea University,
  Seoul, Korea, 136-701}
\author{N. Fornengo}
\affiliation{Dipartimento di Fisica Teorica, Universit\`a di Torino \\
Istituto Nazionale di Fisica Nucleare, Sezione di Torino \\
via P. Giuria 1, I--10125 Torino, Italy}
\author{A. Bottino}
\affiliation{Dipartimento di Fisica Teorica, Universit\`a di Torino \\
Istituto Nazionale di Fisica Nucleare, Sezione di Torino \\
via P. Giuria 1, I--10125 Torino, Italy}

\date{\today}

\begin{abstract}
  We discuss the impact for light neutralinos in an effective Minimal
  Supersymmetric extension of the Standard Model of the recent results
  presented by the CMS and ATLAS Collaborations at the CERN Large
  Hadron Collider for a search of supersymmetry in proton-proton
  collisions at a center-of-mass energy of 7 TeV with an integrated
  luminosity of 35 ${\rm pb}^{-1}$.  We find that, in the specific
  case of light neutralinos, efficiencies for the specific signature
  searched by ATLAS (jets+missing transverse energy and an isolated
  lepton) imply a lower sensitivity compared to CMS (which searches
  for jets +missing transverse energy). Focusing on the CMS bound, if
  squark soft masses of the three families are assumed to be
  degenerate, the combination of the ensuing constraint on squark and
  gluino masses with the experimental limit on the $b \rightarrow s +
  \gamma$ decay imply a lower bound on the neutralino mass $m_{\chi}$
  that can reach the value of 11.9 GeV, depending on the gluino
  mass. On the other hand, when the universality condition among
  squark soft parameters is relaxed, the lower bound on $m_{\chi}$ is
  not constrained by the CMS measurement and then remains at the value
  7.5 GeV derived in previous papers.  

\end{abstract}

\pacs{95.35.+d,11.30.Pb,12.60.Jv,95.30.Cq}

\maketitle


\section{Introduction}
\label{intro}

The CMS and ATLAS Collaborations have presented their results of a
search for supersymmetry (SUSY) in proton-proton collisions at the
CERN Large Hadron Collider (LHC) at a center-of-mass energy of 7 TeV
with an integrated luminosity of 35 ${\rm pb}^{-1}$
\cite{cms,atlas}. The CMS investigation\cite{cms} consists in a search
for events with jets and missing transverse energy, while
ATLAS\cite{atlas} searched for final states containing jets, missing
transverse energy and one isolated electron or muon.  Both signatures
would be significant of processes due to the production in pairs of
squarks and gluinos, subsequently decaying into quarks, gluons, other
standard-model (SM) particles and a neutralino (interpreted as the
lightest supersymmetric particle (LSP)) in a R-parity conserving SUSY
theory.  As reported in Ref.\cite{cms,atlas} in both analyses the data
appear to be consistent with the expected SM backgrounds; thus
constraints are derived on the model parameters in the case of a
minimal supergravity model (mSUGRA, also denoted as CMMS)
\cite{nilles} for the specific standard benchmark with trilinear
coupling $A_0$=0, ratio of vacuum expectation values $\tan\beta$=3,
Higgs--mixing parameter $\mu>0$ in the plane of the universal scalar
and gaugino mass parameters $m_0$--$m_{1/2}$. In
Ref. \cite{cms}constraints are also discussed in terms of two of the
conventional benchmarks within SUGRA models: those denoted by LM1 and
LM0 (or SU4) in the literature
\cite{cms_report,atlas_report,cms_report2}. Though these constraints
depend on the specific sets of the mSUGRA parameters employed in the
phenomenological analysis, the general outcome of
Refs.\cite{cms,atlas} is that the lower bounds on the squark and
gluino masses are sizeably higher as compared to the previous limits
established by the experiments D0 \cite{D0} and CDF \cite {CDF} at the
Tevatron.

In this paper we consider the implications of the results of
Refs.~\cite{cms,atlas} for the supersymmetric scheme discussed in
Refs.~\cite{lowneu,sum,discussing}, {\it i.~e.} for an effective MSSM
scheme at the electroweak scale with the following independent
parameters: $M_1, M_2, M_3, \mu, \tan\beta, m_A, m_{\tilde q},
m_{\tilde l}$ and $A$.  Notations are as follows: $M_1$, $M_2$ and
$M_3$ are the U(1), SU(2) and SU(3) gaugino masses (these parameters
are taken here to be positive), $\mu$ is the Higgs mixing mass
parameter, $\tan\beta$ the ratio of the two Higgs v.e.v.'s, $m_A$ the
mass of the CP-odd neutral Higgs boson, $m_{\tilde q}$ is a squark
soft--mass common to all squarks, $m_{\tilde l}$ is a slepton
soft--mass common to all sleptons, and $A$ is a common dimensionless
trilinear parameter for the third family, $A_{\tilde b} = A_{\tilde t}
\equiv A m_{\tilde q}$ and $A_{\tilde \tau} \equiv A m_{\tilde l}$
(the trilinear parameters for the other families being set equal to
zero).  Since no gaugino-mass unification at a Grand Unified scale is
assumed (at variance with one of the major assumptions in mSUGRA), in
this model the neutralino mass is not bounded by the lower limit
$m_{\chi} \gsim$ 50 GeV that is commonly derived in mSUGRA schemes
from the LEP lower bound on the chargino mass (of about 100 GeV).  In
Refs.\cite{lowneu,sum,discussing} it is shown that, if R-parity is
conserved, a light neutralino ({\it i.~e.} a neutralino with $m_{\chi}
\lsim$ 50 GeV) is a very interesting candidate for cold dark matter
(CDM), due to its relic abundance and its relevance in the
interpretation of current experiments of search for relic particles;
in Refs. \cite{lowneu,sum,discussing} also a lower bound, $m_{\chi}
\gsim$ 7-8 GeV, is obtained from the cosmological upper limit on
CDM. The compatibility of these results with all experimental searches
for direct or indirect evidence of SUSY (prior to the results of
Refs.\cite{cms,atlas}) and with other precision data that set
constraints on possible effects due to supersymmetry is discussed in
detail in Ref.\cite{discussing}.  The SUSY model described above will
hereafter be denoted as Light Neutralino Model (LNM); within this
model, the so-called {\it Scenario $\mathcal{A}$} \cite{discussing}
will be considered in the present analysis.  The main features of this
scenario are: i) $m_A$ must be light, 90 GeV $\leq m_A \lsim
(200-300)$ GeV (90 GeV being the lower bound from LEP searches); ii)
$\tan \beta$ has to be large: $\tan \beta$ = 20--45, iii) the $\tilde
{B} - \tilde H_1^{\circ}$ mixing needs to be sizeable, which in turn
implies small values of $\mu$: $|\mu| \sim (100-200)$ GeV. The purpose
of this paper is to establish the novelties introduced by the outcomes
of the recent CMS and ATLAS investigations on the features of the LNM,
with special emphasis on the aspects concerning the neutralino as a
CDM candidate. For detailed discussions of LNM models, see
Refs. \cite{lowneu,sum} and especially Ref. \cite{discussing}.

First, we recall that the neutralino, defined as the linear
superposition of bino $\tilde B$, wino $\tilde W^{(3)}$ and of the two
Higgsino states $\tilde H_1^{\circ}$, $\tilde H_2^{\circ}$, $\chi
\equiv a_1 \tilde B + a_2 \tilde W^{(3)} + a_3 \tilde H_1^{\circ} +
a_4 \tilde H_2^{\circ}$, of lowest mass $m_{\chi}$, is described
within the minimal supersymmetric extension of the SM only through a
subset of the SUSY model parameters, namely $M_1, M_2, \mu$ and $\tan
\beta$.  The neutral Higgs mass $m_A$ and the slepton mass $m_{\tilde
  l}$ are instead crucial parameters intervening in the
neutralino-nucleon scattering and in the neutralino pair-annihilation
processes (and then also in the neutralino relic abundance)
\cite{lowneu,sum,discussing}.  The three remaining parameters
characterizing the LNM: $M_3$, $m_{\tilde q}$ and $A$, enter into
play, when the large host of experimental results that constrain
supersymmetry are implemented into the model \cite{discussing}.  This
experimental information is derived from : 1) the searches at
accelerators for Higgs bosons and supersymmetric charged particles
(sleptons and charginos at LEP, squarks and gluinos at hadron
colliders); 2) the B-meson rare decays at the Tevatron and the
B-factories; 3) the muon anomalous magnetic moment; 4) the $b
\rightarrow s \gamma$ decay.  One further crucial requirement which
guarantees that the neutralino can be interpreted as a relic particle
in the Universe is that its relic abundance satisfies the cosmological
bound $\Omega_{\chi} h^2 \leq (\Omega_{CDM} h^2)_{\rm max} \simeq$
0.12.  All these data set significant constraints on the model
parameters and also entail sizable correlations among some of them. In
particular, various constraints and correlations involving the SUSY
parameters follow from the loop correction terms, due to
supersymmetry, that can affect the physical quantities involved in the
items (2-4) above.

\section{Correlations between $m_{\tilde q}$ and other SUSY parameters within the LNM}
\label{corr}

\begin{figure}[t]
  \includegraphics[width=0.9\columnwidth,clip=false,bb=33 59 515
  505]{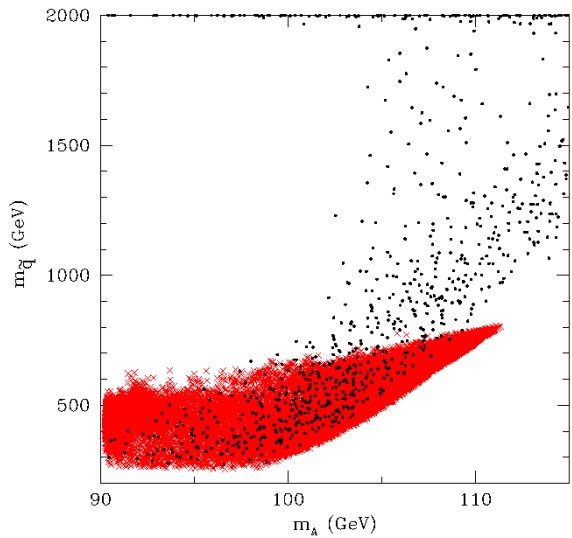}
  \caption{Scatter plot of the light neutralino population shown in
    the plane $m_{A}-m_{\tilde{q}}$. For (black) dots the $b
    \rightarrow s + \gamma$ constraint is not implemented, while for
    (red) crosses the constraint is applied.}
\label{fig:scat_ma_m_sq}
\end{figure}

One of the most important constraints among those mentioned above is
the one established by the branching ratio of the $b \rightarrow s +
\gamma$ decay process. Indeed, in the LNM this branching ratio lies in
its experimental range if the contribution of a loop diagram with a
charged Higgs and the top quark is compensated by the contribution of
a loop diagram with a chargino and a top squark \cite{garisto}. Since
our LNM (in {\it Scenario $\mathcal{A}$}) entails both a light charged
Higgs (of mass $m^2_{H^{\pm}} \simeq m^2_A + m^2_W$) and a light
chargino (of mass $m_{{\chi}^{\pm}} \sim \mu \sim$ 100-200 GeV), also
$m_{\tilde q}$ has to be not too heavy. A strong correlation implied
by the $b \rightarrow s + \gamma$ decay process between $m_A$ (through
$m_{H^{\pm}}$) and $m_{\tilde q}$ is shown in
Fig.~\ref{fig:scat_ma_m_sq}, where a scatter plot for a light
neutralino population is represented by (black) dots when the $b
\rightarrow s + \gamma$ constraint is not implemented, and by (red)
crosses when this constraint is applied. In this second case it turns
out that: i) $m_{\tilde q}$ and $m_A$ are rather strongly correlated,
ii) the squark mass is limited by the upper bound $m_{\tilde q} \lsim$
800 GeV. Notice that the variation in the density of points of
Fig. \ref{fig:scat_ma_m_sq} is just due to a different sampling of the
regions of interest in the parameter space. The relaxation of the
$b\rightarrow s \gamma$ constraint is only considered here in
connection with Fig. \ref{fig:scat_ma_m_sq}, for illustration
purposes. This constraint is implemented in all our further
discussions and results.

\begin{figure}[t]
  \includegraphics[width=0.9\columnwidth,clip=true,bb=33 59 515 505]{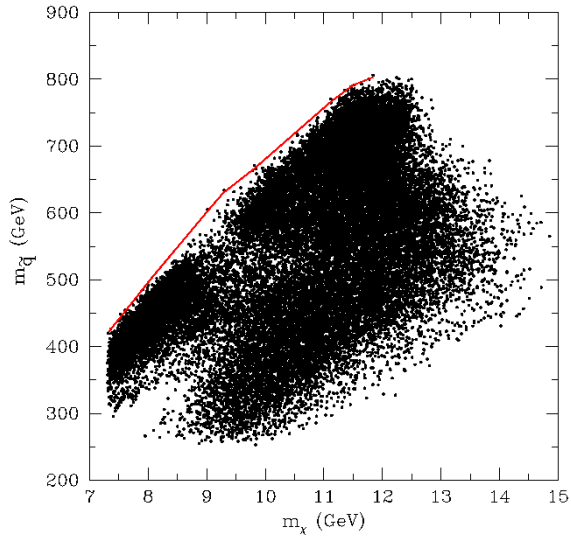}
  \caption{Scatter plot of the light neutralino population shown in
    the plane $m_{\chi}-m_{\tilde{q}}$. The (red) line represents an
    interpolation of the lower boundary on $m_{\chi}$ as a function of
    $m_{\tilde{q}}$.}
\label{fig:scat_mchi_m_sq}
\end{figure}

Since the lower bound on the neutralino mass, implied by the
cosmological bound $\Omega_{\chi} h^2 \leq (\Omega_{CDM} h^2)_{\rm
  max}$, increases as $m^2_A$ \cite{discussing}, the correlation
between $m_{\tilde q}$ and $m_A$ entails also a correlation between
$m_{\tilde q}$ and $m_{\chi}$, as displayed in
Fig.\ref{fig:scat_mchi_m_sq}.

These correlations imply that a lower bound on $m_{\tilde q}$, derived
from accelerator measurements could potentially have the consequence
of increasing the lower bound on $m_{\chi}$, as compared to the one of
about 7--8 GeV, previously established within the LNM \cite{discussing}.
Thus, it is important to establish which lower limit on $m_{\tilde q}$
can be actually derived from the CMS and ATLAS results
\cite{cms,atlas}.

\begin{figure}[t]
  \includegraphics[width=0.9\columnwidth,clip=true,bb=33 59 515 505]{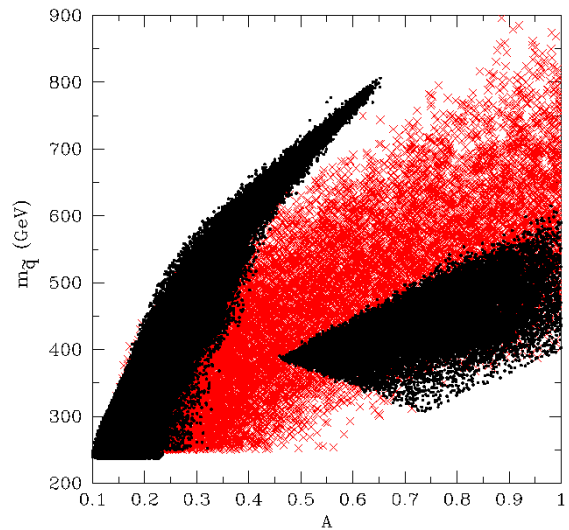}
  \caption{Scatter plot of the light neutralino population shown in
    the plane $A-m_{\tilde{q}}$. (Black) dots show configurations
    for which all constraints are applied, while for
    (red) crosses the bound from $B \rightarrow \tau \nu$ measurements
    \protect\cite{discussing} is not implemented.}
\label{fig:scat_atrilinear_m_sq}
\end{figure}

Before we come to an analysis of this point, let us just remark that a
loop involving the chargino and the stop, as the one relevant for the
$b \rightarrow s + \gamma$, is also responsible for a potentially
sizable SUSY contribution to the branching ratio for the decay $B_s
\rightarrow {\mu}^{+} + {\mu}^{-}$. Indeed, this loop correction
behaves as $\tan^6 \beta$ \cite{bobeth}, thus, at large $\tan \beta$,
it can overshoot the experimental upper bound: $BR(B_s \rightarrow
{\mu}^{+} {\mu}^{-}) < 5.8 \times 10^{-8}$ \cite{cdf_mumu}.  This can
actually occur in SUGRA models, with the effect of constraining the
neutralino phenomenology drastically \cite{nath}.  In
Ref.~\cite{discussing} it is shown that in the LNM this is not the
case, since: (a) the chargino intervening in the relevant loop is
light, and (b) the splitting in the top mass eigenstates can be small
(a condition that is met whenever: $|A| \ll m_{\tilde q}/m_t$).  This
last requirement is exemplified by the lower frontier of the scatter
plot of Fig.~\ref{fig:scat_atrilinear_m_sq} displaying the correlation
between $A$ and $m_{\tilde q}$.  In this figure the upper bound on
$m_{\tilde{q}}$ is due, as already mentioned, to the bound on the $b
\rightarrow s + \gamma$ decay.  The point we wish to stress here is
that, as shown in the numerical analysis of Ref.~\cite{discussing},
the constraint imposed by the branching ratio for the decay $B_s
\rightarrow {\mu}^{+} + {\mu}^{-}$ is compatible with the constraints
due to the branching ratio of the $b \rightarrow s + \gamma$ decay
process, a feature which is not trivial, due to the different role
played by the parameter $m_{\tilde q}$ in the two processes.  However,
it is clear from Fig. \ref{fig:scat_atrilinear_m_sq} that as the
squark soft mass parameter $m_{\tilde{q}}$ gets close to its upper
bound, the interplay of the two constraints entails a growing tuning
of the $A$ trilinear coupling for the highest values
of $m_{\tilde{q}}$. In the same figure (black) dots show
configurations for which all constraints are applied, while for (red)
crosses the bound from $B \rightarrow \tau \nu$ measurements is not
implemented. As discussed in Ref. \cite{discussing}, this latter bound
is somewhat less robust than other constraints, due to the
uncertainties affecting both theoretical estimates and experimental
determinations related to B--meson decays. As can be seen from
Fig. \ref{fig:scat_atrilinear_m_sq}, when the $B \rightarrow \tau \nu$
constraint is not implemented the tuning affecting the trilinear
coupling is eased and the upper bound on $m_{\tilde{q}}$ weakened.

\section{Lower limit to $m_{\tilde q}$ implied by the CMS and ATLAS
  results within the LNM}

After appropriate cuts to reject the background and to reduce the
probability of jet mismeasurements, the CMS search for events with
jets and missing transverse energy derived an upper bound $N_{\rm
  max}^{\rm CMS}$= 13 events at 95\% confidence level in the signal
region for an integrated luminosity ${\cal L}= 35$ pb $^{-1}$. This
upper bound is related to the total SUSY production cross section
$\sigma$ by the relation $N_{\rm max}=\epsilon\times {\cal L}\times
\sigma$, where $\epsilon$ is the total efficiency due to selection
cuts.  In order to estimate $\epsilon_{\rm CMS}$ for the CMS signature
we have simulated a few LNM benchmarks on the low--$m_{\chi}$ boundary
shown in Fig. \ref{fig:scat_mchi_m_sq} using ISAJET \cite{isajet},
applying the same kinematic cuts as described in Ref. \cite{cms}. In
this way we obtained the range $0.07\lsim\epsilon_{\rm CMS}\lsim 0.2$
for the total efficiency, that can be used to convert $N_{\rm
  max}^{\rm CMS}$ into an upper bound $\sigma_{\rm CMS}^{\rm max}$ on
the cross section, with $1.86\,{\rm pb}<\sigma_{\rm CMS}^{\rm
  max}<5.31\,{\rm pb}$.  On the other hand, the ATLAS collaboration
searched for jets+missing transverse energy and one isolated electron
or muon, and derived an upper bound $N_{\rm max}^{\rm ATLAS}$= 2.2
events at 95\% confidence level in the electron signal region (with a
similar result in the muon channel) for the same integrated luminosity
of CMS. Following the same procedure used for CMS and for the same LNM
benchmarks, we estimated $\epsilon_{\rm ATLAS}$ for the ATLAS
signature applying the same kinematic cuts as described in
Ref. \cite{atlas}. In this way we found the range $2\times
10^{-4}\lsim\epsilon_{\rm ATLAS}\lsim 5\times 10^{-3}$, that when
converted into an upper bound on the cross section $\sigma_{\rm
  ATLAS}^{\rm max}$ implies $12.6\,{\rm pb}<\sigma_{\rm ATLAS}^{\rm
  max}<314.3\,{\rm pb}$. Since $\sigma_{\rm ATLAS}^{\rm max}\gg
\sigma_{\rm CMS}^{\rm max}$ we conclude that, within the LNM scenario,
the CMS analysis is significantly more sensitive than that from
ATLAS\footnote{We find that the particular suppression of
  $\epsilon_{\rm ATLAS}$ is due to the cut on the angle between the
  missing transverse momentum vector and the jets, applied by ATLAS to
  reduce the probability of jet mismeasurement \protect\cite{atlas}.}.
As a consequence of the above discussion, in the following we will
concentrate only on the discussion of the CMS bound.

\begin{figure}[t]
  \includegraphics[width=1.1\columnwidth,clip=true,bb=0 185 590 651]{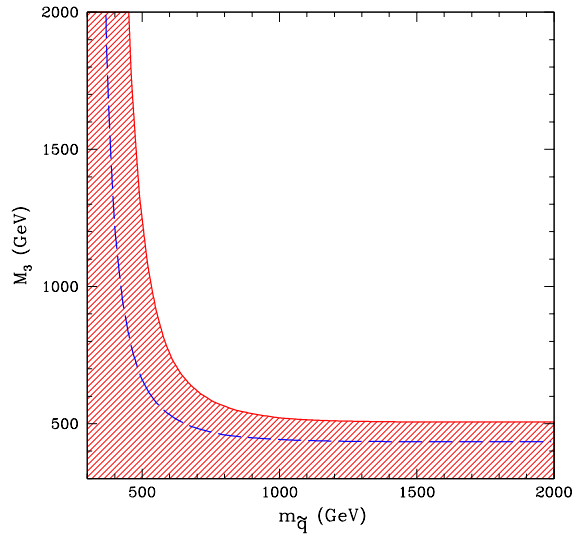}
  \caption{Shaded area representing the region in the
    $m_{\tilde{q}}$-$M_3$ parameter space where the total SUSY
    production cross section at the LHC with $\sqrt{s}=7$ TeV is
    larger than 1.86 pb, corresponding to the CMS upper bound of 13
    SUSY events \protect\cite{cms}, and assuming an average total
    efficiency due to kinematic cuts equal to 0.2 for the LNM
    scenario. The (blue) dashed line represents the contour plot for
    $\sigma=\sigma_{\rm CMS}$= 5.31 pb, the value
    corresponding to the upper bound on the cross section when
    $\epsilon_{\rm CMS}=0.07$ (see text).}
\label{fig:cross_section_contour}
\end{figure}

In Fig.\ref{fig:cross_section_contour} the solid (red) line shows the
contour plot for $\sigma=\sigma^{\rm CMS}=1.86$ pb, while the dashed
(blue) one represents the corresponding curve for $\sigma=\sigma^{\rm
  CMS}=5.31$ pb; we have calculated the total SUSY production cross
section for the process $p+p\rightarrow$ gluinos, squarks as a
function of the squark mass $m_{\rm squark}\simeq m_{\tilde{q}}$ and
the gluino mass $M_3$ using PROSPINO \cite{prospino} with CTEQ-TEA
CT10 Parton Distribution Functions \cite{ct10}. The shaded area below
the (red) solid line would be excluded adopting $\epsilon_{\rm
  CMS}=0.2$ and represents the maximal impact of the CMS measurement
on the LNM parameter space.  It is important here to point out that,
at variance with the SUGRA scenario, within the LNM model the gluino
mass $M_3$ is not related to the other gaugino masses, and in
particular to $m_{\chi}\simeq M_1$ by GUT relations. Moreover, $M_3$
enters in the calculation of observables for the relic neutralino only
at the loop level (through radiative corrections of Higgs couplings
\cite{qcd_corr}) so that within the LNM $M_3$ is very weakly
correlated to the other parameters. This implies that within the LNM
the absolute lower bound $m_{\tilde{q}}\gsim$ 450 (370) GeV can be obtained
from the contour plot of Fig. \ref{fig:cross_section_contour} by
taking the limit $M_3\rightarrow \infty$ and for $\epsilon_{\rm
  CMS}=0.2 (0.07)$.

\section{Lower limit to $m_{\chi}$ implied by the CMS results within
  the LNM for degenerate squark soft masses}

\begin{figure}[t]
  \includegraphics[width=1.1\columnwidth,clip=true,bb=0 185 590 651]{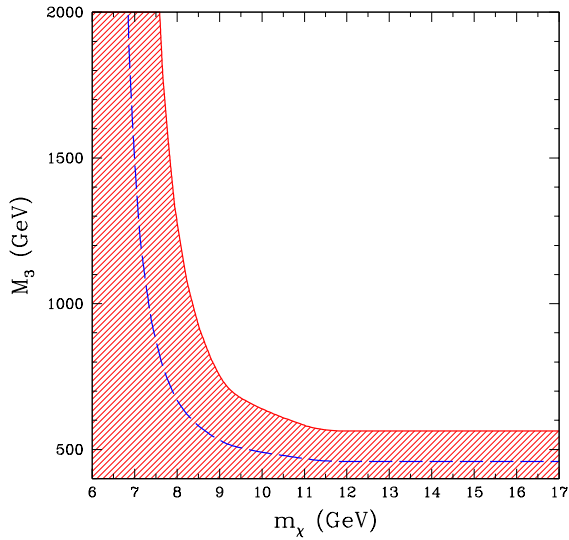}
  \caption{Lower bound on the neutralino mass $m_{\chi}$ as a function
    of the gluino mass $M_3$, that can be derived from the CMS data
    \protect\cite{cms} when soft mass parameters of squarks of the
    three families are assumed to be degenerate in the LNM. The solid
    (red) line is obtained adopting the efficiency $\epsilon_{\rm
      CMS}=0.2$ and the dashed (blue) one corresponds to the case
    $\epsilon_{\rm CMS}=0.07$. }
\label{fig:mchi_vs_m3}
\end{figure}

As already mentioned before, within the LNM the $m_{\tilde{q}}$
parameter is correlated to the neutralino mass $m_{\chi}$, as shown by
the scatter plot of Fig.\ref{fig:scat_mchi_m_sq}. As a consequence,
the lower bound on $m_{\tilde{q}}$ discussed in the previous Section
can be converted into a lower bound on $m_{\chi}$. This is shown as a
function of $M_3$ in Fig.\ref{fig:mchi_vs_m3}, where the solid (red)
line corresponds to $\epsilon_{\rm CMS}=0.2$ and the dashed (blue) one
to $\epsilon_{\rm CMS}=0.07$.  In both cases the boundary shown in
Fig. \ref{fig:scat_mchi_m_sq} has been used to convert the bound on
$m_{\tilde{q}}$ into a limit on $m_{\chi}$.  Notice that, assuming
degenerate soft squark masses, in the LNM the CMS limit can be
combined to the upper bound $m_{\tilde{q}}<$ 800 GeV obtained from the
$b \rightarrow s + \gamma$ decay process to get the absolute limit
$M_3>$ 560 (460) GeV for $\epsilon_{\rm CMS}=0.2(0.07)$. For this
reason the bound of Fig. \ref{fig:mchi_vs_m3} becomes a flat line for
$m_{\chi}\gsim$ 11.8 (11.9) GeV. From this figure we also notice that
the absolute lower bound on $m_{\chi}$ is 7.6 (6.8) GeV. This bound is
increased to 11.8 (11.9) GeV when the gluino mass is close to its
lower limit of 560 (460) GeV. In Fig. \ref{fig:mchi_vs_m3} the shaded
area below the (red) solid line would be excluded adopting
$\epsilon_{\rm CMS}=0.2$ and represents the maximal impact of the CMS
measurement on the LNM parameter space.

\begin{figure}
  \includegraphics[width=0.9\columnwidth,clip=true,bb=33 59 515 505]{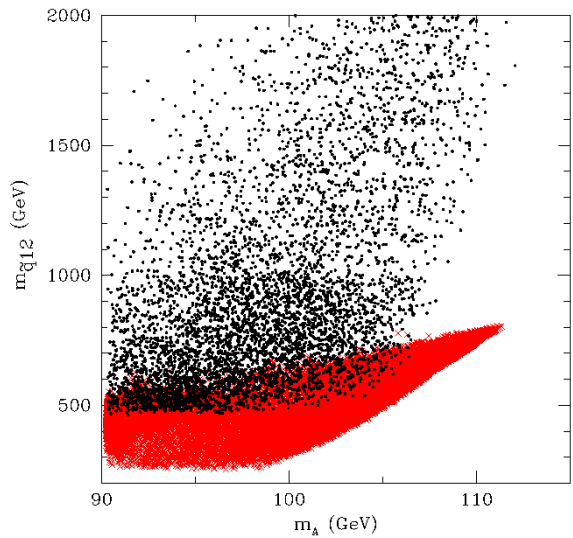}
  \caption{Scatter plot of the light neutralino population in the
    plane $m_A$-- $m_{squark}$. For (red) crosses the squark
    soft--mass parameters are assumed to be degenerate, $m_{{\tilde
        q}_{12}}$=$m_{{\tilde t}}\equiv m_{\tilde{q}}$, while for
    (black) dots $m_{{\tilde q}_{12}}$ and $m_{{\tilde t}}$ are
    allowed to float independently. }
\label{fig:ma_vs_m_sq_split}
\end{figure}

\section{Extension of the LNM by removing the degeneracy in  $m_{\tilde q}$}

According to the previous derivations we can conclude that, within the
LNM described in terms of the eight SUSY parameters, the squark-mass
parameter has to stay in the range (370) 450 GeV $\lsim m_{\tilde q} \lsim$
800 GeV, with the further feature that in the high side of this range
the model requires some fine-tuning. These properties are strictly
related to the choice we have made before of taking a single soft mass
parameter $m_{\tilde q}$ for all squarks; a choice originally taken to
keep the number of SUSY parameters as low as possible.  We consider
here a minimal extension of the previous LNM, by removing this
degeneracy in $m_{\tilde q}$. A natural (SUGRA-inspired) hierarchy
among the soft squark masses might consists in introducing a common
soft mass for the first two families, $m_{{\tilde q}_{12}}$, larger
than the soft mass parameter for the third family, $m_{{\tilde t}}$.
We expect this splitting to reduce the fine tuning discussed in the
previous Sections because LHC physics is mainly sensitive to squarks
of the first two families (which correspond to the flavors more
abundant in colliding protons), while the dominant contribution to the
$b \rightarrow s + \gamma$ decay is driven by the large Yukawa
coupling of the top squark. This is confirmed by the scatter plot in
Fig. \ref{fig:ma_vs_m_sq_split}, where (red) crosses represent the
same configurations shown in Fig. \ref{fig:scat_ma_m_sq} with
$m_{{\tilde q}_{12}}$=$m_{{\tilde t}}\equiv m_{\tilde{q}}$, while
(black) dots show configurations where $m_{\tilde{q}_{12}}$ and
$m_{{\tilde t}}$ are allowed to float independently. In this latter
case the $m_{\tilde{q}_{12}}$ parameter is no longer constrained from
above for all values of $m_A$.  As a consequence, in this case
$m_{\chi}$ is no longer constrained by the CMS measurement.

An analysis of the capability of the LHC in exploring SUSY regions
where the first generation squarks are very heavy compared to the
other superpartners is performed in Ref.~\cite{fan}, but for models
different from LNM.

\section{Conclusions}

In this paper we have discussed the impact for light neutralinos in an
effective Minimal supersymmetric extension of the Standard Model of
the recent results presented by the CMS and ATLAS Collaborations at
the CERN Large Hadron Collider for a search of supersymmetry in
proton-proton collisions at a center-of-mass energy of 7 TeV with an
integrated luminosity of 35 ${\rm pb}^{-1}$.  Within the LNM model we
found that CMS is significantly more sensitive than ATLAS, due to the
different signatures searched by the two experiments.  In particular,
we estimated a detection efficiency at CMS $0.07\lsim\epsilon_{\rm
  CMS}\lsim 0.2$ after kinematic cuts, corresponding to an upper bound
for the total SUSY production cross section that varies from
$1.86\,{\rm pb}$ to $5.31\,{\rm pb}$ . Taking the limit $M_3\gg
m_{\tilde{q}}$ this implies an absolute lower bound of 450 (370) GeV
for the squark mass when $\epsilon_{\rm CMS}$=0.2(0.07). If squark
soft masses of the three families are assumed to be degenerate, we
found that the combination of the CMS bound on the squark mass with
the experimental constraints on the $b \rightarrow s + \gamma$ and the
$B_s \rightarrow {\mu}^{+} + {\mu}^{-}$ decays entail some tuning of
the $A$ trilinear coupling at high values of
$m_{\tilde{q}}$. Moreover, when combining the CMS bound to the $b
\rightarrow s + \gamma$ constraint the lower bound on the neutralino
mass $m_{\chi}$ varies between 6.8 and 11.9 GeV, depending on the
gluino mass. On the other hand, if the universality condition among
squark soft parameters is relaxed the CMS measurement implies no
constraint on the lower limit on $m_{\chi}$, that remains at the value
7.5 GeV as derived in Ref. \cite{discussing}.

\acknowledgments

A.B. and N.F. acknowledge Research Grants funded jointly by Ministero
dell'Istruzione, dell'Universit\`a e della Ricerca (MIUR), by
Universit\`a di Torino and by Istituto Nazionale di Fisica Nucleare
within the {\sl Astroparticle Physics Project} (MIUR contract number:
PRIN 2008NR3EBK; INFN grant code: FA51). S.S. acknowledges support by
NRF with CQUEST grant 2005-0049049 and by the Sogang Research Grant
2010. N.F. acknowledges support of the spanish MICINN Consolider
Ingenio 2010 Programme under grant MULTIDARK CSD2009- 00064.
S.C. acknowledges support from the Korean National Research Foundation
NRF-2010-0015467.

\end{document}